\documentclass[10pt,journal,compsoc]{IEEEtran}
\def\IEEEcompsocdiamondline{\vspace{-0.7cm}}

\usepackage{amsmath,amssymb,amsfonts}
\usepackage{algorithmic}
\usepackage{graphicx}
\usepackage{textcomp}
\usepackage{xcolor}
\usepackage[hyphens]{url}
\usepackage{booktabs}
\usepackage{pifont}
\usepackage{siunitx}
\usepackage{amsmath}
\usepackage{algorithm2e}
\usepackage{multirow}
\usepackage{fancyhdr}
\usepackage{float}
\usepackage{listings}
\usepackage{balance}
\usepackage{caption}
\usepackage{stfloats}
\usepackage[hidelinks]{hyperref}
\usepackage{cite}
  
\usepackage[sort&compress, numbers]{natbib}
\usepackage{enumitem}

\usepackage{titlesec}
\titlespacing*{\section}{0pt}{0.5\baselineskip}{0.3\baselineskip}
\titlespacing{\subsection}{0pt}{0.5\baselineskip}{0.3\baselineskip}
\titlespacing{\subsubsection}{0pt}{0.5\baselineskip}{0.3\baselineskip}
\definecolor{moegi}{rgb}{0.357, 0.537, 0.188}

\definecolor{codekeyword}{HTML}{D32F2F}
\definecolor{codestring}{HTML}{4CAF50}
\definecolor{codecomment}{HTML}{9E9E9E}

\makeatletter
\newcommand{\srcsize}{\@setfontsize{\srcsize}{7pt}{7pt}}
\makeatother

\lstdefinestyle{cppstyle}{
    numbers=left,
    language=C++,
    basicstyle=\ttfamily\srcsize,
    commentstyle=\color{codecomment},
    keywordstyle=\color{codekeyword}\bfseries,
    stringstyle=\color{codestring},
    frame=single,
    rulecolor=\color{black},
    xleftmargin=1.6em,
    aboveskip=0.2em,
    belowskip=1em, %
    showstringspaces=false,
    breaklines=true,
    breakatwhitespace=true,
    tabsize=2,
    morekeywords={constexpr,decltype,nullptr,noexcept,override},
}

\DeclareCaptionFormat{listing}{\par\vskip1pt#1#2#3}
\definecolor{codekeyword}{HTML}{D32F2F}
\definecolor{codestring}{HTML}{4CAF50}
\definecolor{codecomment}{HTML}{9E9E9E}

\lstdefinestyle{yamlstyle}{
    language=YAML,
    basicstyle=\ttfamily\footnotesize,
    commentstyle=\color{codecomment},
    keywordstyle=\color{codekeyword}\bfseries,
    stringstyle=\color{codestring},
    frame=single,
    rulecolor=\color{black},
    xleftmargin=1em,
    aboveskip=1em,
    belowskip=1em,
    showstringspaces=false,
    breaklines=true,
    breakatwhitespace=true,
    tabsize=2,
    morekeywords={true,false,yes,no},
    numbers=left,
}

\newcommand{\secref}[1]{Section~\ref{#1}}
\newcommand{\figref}[1]{Figure~\ref{#1}}

\usepackage{setspace}
\usepackage{tikz}
\newcommand*\circled[1]{\tikz[baseline=(char.base)]{\node[shape=circle,fill,inner sep=.8pt] (char) {\textcolor{white}{#1}};}}
\newcommand*\hcircled[1]{\tikz[baseline=(char.base)]{\node[shape=circle, draw=black, fill=white, inner sep=.8pt] (char) {\textcolor{black}{#1}};}}

\usepackage{fancyhdr}
\usepackage{glossaries}

\ifCLASSOPTIONcompsoc
\else
  \usepackage{cite}
\fi

\ifCLASSINFOpdf
\else
\fi

\begin{document}
\title{Ramulator 2.0: A Modern, Modular, and Extensible DRAM Simulator}

\author{Haocong Luo,
        {Yahya Can Tu\u{g}rul},
        {F. Nisa Bostancı},
        {Ataberk Olgun},
        {A. Giray Ya\u{g}l{\i}k\c{c}{\i}},
        and~Onur Mutlu
        \vspace{-1em}
}

\IEEEtitleabstractindextext{%
\begin{abstract}
We present Ramulator 2.0, a highly modular and extensible DRAM simulator that enables rapid and agile {implementation} and {evaluation} of design changes in the memory controller and DRAM to meet the increasing research effort in improving the performance, security, and reliability of memory systems. Ramulator 2.0 abstracts and models key components in a DRAM-based memory system and their interactions into shared \emph{interfaces} and independent \emph{implementations}. {Doing so enables easy modification and extension of the modeled functions of the memory controller and DRAM in Ramulator 2.0.} The DRAM specification syntax of Ramulator 2.0 is concise and human-readable, facilitating easy modifications and extensions. Ramulator 2.0 implements a library of reusable templated lambda functions to model the functionalities of DRAM commands to simplify the implementation of new DRAM standards, including DDR5, LPDDR5, HBM3, and GDDR6. We showcase Ramulator 2.0's modularity and extensibility by implementing and evaluating a wide variety of RowHammer mitigation techniques that {require} \emph{different} memory controller design changes. {These techniques are added modularly as separate implementations} \emph{without} changing {\emph{any}} code in the baseline memory controller implementation. Ramulator 2.0 is {rigorously} validated and maintains a fast simulation speed compared to existing cycle-accurate DRAM simulators. Ramulator 2.0 is open-sourced under the permissive MIT license {at \url{https://github.com/CMU-SAFARI/ramulator2}}.

\end{abstract}

\vspace{-1.5em}
}

\maketitle

\IEEEdisplaynontitleabstractindextext

\IEEEpeerreviewmaketitle

\section{Introduction}\label{sec:introduction}

Cycle-accurate DRAM simulators {enable modeling and evaluation of} detailed operations in the memory controller and the DRAM device. In recent years, growing research {and design} efforts in improving the performance, security, and reliability of DRAM-based memory systems require a cycle-accurate simulator that facilitates rapid and agile {implementation} and {evaluation} of intrusive design changes (i.e., {modification of functionalities of the simulated system as opposed to simple parameter changes}) in the memory controller and {DRAM}. Unfortunately, existing cycle-accurate DRAM simulators are not modular and extensible {\emph{enough}} to meet such a requirement.

We identify two key issues in the design and implementation of existing cycle-accurate DRAM simulators. First, {{they} do \emph{not} model key components of a DRAM-based memory system in a {\emph{fundamentally modular}} way, making it difficult to implement and maintain different intrusive design changes.}
For example, USIMM~\cite{usimm} does not separate the DRAM {specification} from the memory controller. {Similarly,} the templated implementations of the DRAM specifications in Ramulator~\cite{kim2016ramulator} (referred to as Ramulator 1.0 in this paper) cause undesired coupling between the DRAM specification and the memory controller.

Second, existing simulators {do} \emph{not} implement DRAM specifications in a concise and intuitive way, making it difficult to add new DRAM commands and define new timing constraints.
For example, both DRAMsim2~\cite{dramsim2} and DRAMsim3~\cite{dramsim3} {implement} a single DRAM device model that aggregates all the DRAM specifications from all supported DRAM standards in a single C++ class. Ramulator 1.0's DRAM specifications are based on low-level and verbose C++ syntax (e.g., it uses eight {full} lines of C++ code just to specify {solely} a single \texttt{tCCD\_{L}} timing constraint in DDR4~\cite{jedec2017ddr4}). 

To address these issues, we present Ramulator 2.0~\cite{ramulator2github}, a successor to Ramulator 1.0~\cite{kim2016ramulator} that provides an easy-to-use, modular, and extensible software infrastructure for rapid and agile implementation and evaluation of DRAM-related research {and design ideas}. Ramulator 2.0 has two distinguishing features. First, it implements a modular and extensible code framework by identifying and modeling the key components in a DRAM-based memory system into separate \emph{interfaces} and \emph{implementations}. With this framework, different design changes (e.g., different address mapping schemes, request scheduling policies, {new DRAM standards}, RowHammer mitigations) can be implemented as {\emph{independent}} implementations that share the {\emph{same}} interface, {enabling easy modification and extension of Ramulator 2.0}.

Second, to facilitate easy modification of DRAM specifications {(e.g., DRAM organization, commands, timing constraints)}, Ramulator 2.0 implements concise and human-readable definitions of DRAM specifications on top of the lookup table based hierarchical DRAM device model in Ramulator 1.0. {Ramulator 2.0's} DRAM specifications 1) are defined with simple string literals, 2) leverage permutations of different DRAM commands to concisely define timing constraints, and 3) use a library of templated lambda functions that are \emph{reusable} across different DRAM standards to define the functionalities of DRAM commands (e.g., the same RFM command implementation can be {(and is)} used by DDR5~\cite{jedec2020ddr5}, LPDDR5~\cite{jedec2020lpddr5}, and GDDR6~\cite{jedec2021gddr6}, {HBM3~\cite{jedec2023hbm3}}). These improvements are implemented with the new features of C++20~\cite{cpp20} (e.g., constant-evaluated immediate functions), enabling significant duplicate-code reduction and easy modification and extension of the modeled DRAM device's functionalities \emph{without} sacrificing simulation speed. 

We showcase the modularity and extensibility of Ramulator 2.0 by implementing and evaluating a {variety} of RowHammer mitigation {techniques} (PARA~\cite{kim2014flipping}, TWiCe~\cite{lee2019twice}, Graphene~\cite{park2020graphene}, Hydra~\cite{qureshi2022hydra}, Randomized Row-Swap (RRS)~\cite{saileshwar2022randomized}, and an ideal refresh-based mitigation~\cite{kim2020revisiting}) {that} require \emph{different} additional functionalities in the memory controller. These RowHammer mitigations {plug} themselves into the \emph{same} baseline memory controller implementation \emph{without} changing the memory controller's code{, which was not possible in Ramulator 1.0~\cite{kim2016ramulator} and is not possible in any other DRAM simulator {we are aware of}~\cite{dramsim2,dramsim3,usimm}.} 

{The key features and contributions of Ramulator 2.0 {are}:}
\begin{itemize}[leftmargin=*]
\item Ramulator 2.0 is a modular and extensible DRAM simulator written in C++20 that enables rapid and agile implementation and evaluation of design changes in the memory system. Ramulator 2.0 can either work as a standalone simulator, or be used as a memory system library by a system simulator (e.g., gem5~\cite{gem5}{, zsim~\cite{sanchez2013zsim}}).
\item We showcase the modularity and extensibility of Ramulator 2.0 by implementing and evaluating six different RowHammer mitigation {techniques} as plugins to a single {unmodified} memory controller implementation.
\item Ramulator 2.0 implements a wide range of new DRAM standards, including DDR5~\cite{jedec2020ddr5}, LPDDR5~\cite{jedec2020lpddr5}, HBM3~\cite{jedec2023hbm3}, and GDDR6~\cite{jedec2021gddr6} {(as well as old ones, e.g., DDR3~\cite{jedec2012ddr3}, DDR4~\cite{jedec2017ddr4}, HBM(2)~\cite{jedec2021hbm})}.
\item Ramulator 2.0 is {rigorously} validated and maintains a fast simulation speed compared to existing cycle-accurate DRAM simulators.
\item We open-source Ramulator 2.0~\cite{ramulator2github} under the permissive MIT license to facilitate and encourage open research {and agile implementation of new ideas} in memory systems. {We also integrate it with gem5~\cite{gem5}.}
\end{itemize}

\begin{figure*}[b]
\vspace{-1.3em}
\centering
  \includegraphics[width=\textwidth]{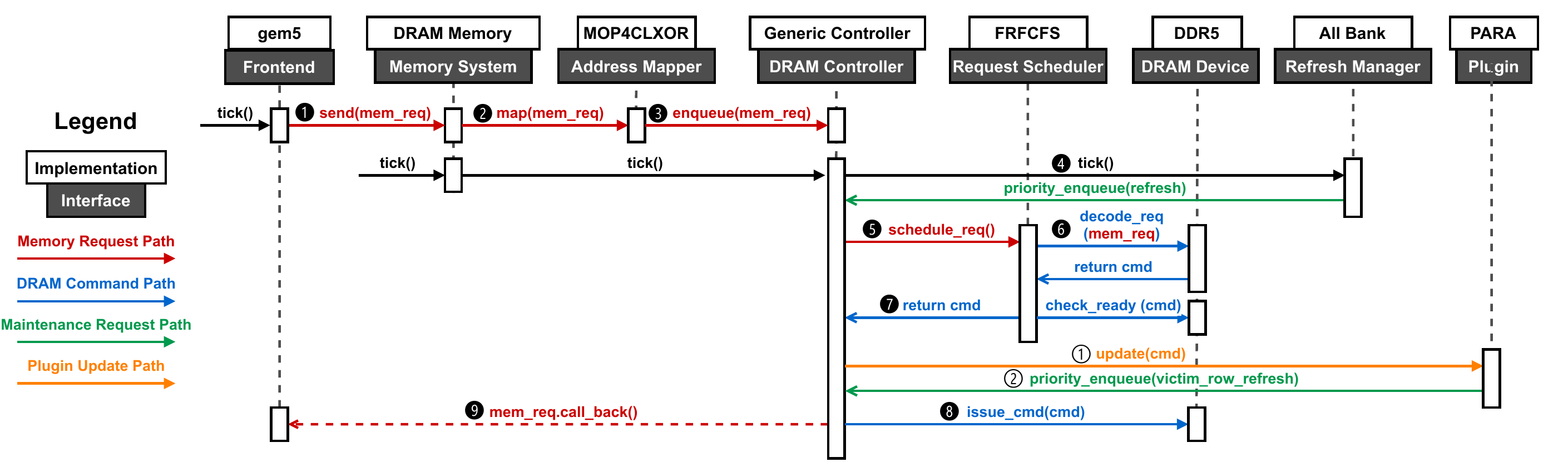}
  \caption{High-level software architecture of Ramulator 2.0 using an example {DDR5 system} configuration}
  \label{fig:ramulator_architecture}

\end{figure*}

\section{Ramulator 2.0 Design Features}
We walk through the two key design features of Ramulator 2.0 that {enable} rapid and agile implementation of design changes in the memory system. \secref{sec:ifce_impl} introduces the high-level software architecture of Ramulator 2.0 based on the key concepts of \emph{interface(s)} and \emph{implementation(s)}. \secref{sec:ctrl_plugin} {provides} a deeper look into the modularity and extensibility enabled by Ramulator 2.0 by showcasing how different RowHammer mitigations can all be implemented as \emph{plugins} of the same baseline {unmodified} memory controller implementation. \secref{sec:dram_model} introduces the concise and human-readable DRAM specification syntax of Ramulator 2.0 that facilitates easy modification and extension {of} the {functionality} of the DRAM device.

\subsection{Modular and Extensible Software Architecture}
\label{sec:ifce_impl}
Ramulator 2.0 models all components in a DRAM-based memory system with two fundamental concepts, \emph{Interface} and \emph{Implementation}, to achieve high modularity and extensibility. An interface is an abstract C++ class defined in a \texttt{.h} header file that models the common high-level functionality of a component as seen by other components in the system. An implementation is a concrete C++ class defined in a \texttt{.cpp} file that inherits from an interface, modeling the actual behavior of a component. Components interact with each other through pointers to each other's interfaces stored in the implementations. With such a design, the functionality of a component can be easily changed by instantiating a different implementation for the same interface, involving {\emph{no}} changes in the code of unrelated components.

Figure~\ref{fig:ramulator_architecture} shows the high-level software architecture of Ramulator 2.0 with the key interfaces we identify in a DRAM-based memory system (dark boxes) and their typical implementations (light boxes) when modeling a {DDR5} system with RowHammer mitigation. The arrows illustrate the {relationships} among different components in the simulated system (i.e., how they call each other's interface functions). We highlight the memory request path with red arrows, DRAM command path with blue arrows, and DRAM maintenance requests (e.g., refreshes) with green arrows. A typical execution of the simulation is as follows: First, memory requests are sent \circled{1} from the frontend (either parsed from traces or generated by {another simulator, e.g., gem5~\cite{gem5}}) to the memory system, where the memory addresses are mapped \circled{2} to the DRAM organization through the address mapper. Then, the requests are enqueued \circled{3} in the request buffers of the DRAM controller. The DRAM controller is responsible for 1) ticking the refresh manager \circled{4}, which could enqueue high-priority maintenance requests ({e.g.}, {refreshes}) back to the controller, 2) querying the request scheduler \circled{5}, which in turn consults the DRAM device model \circled{6} to {decode} the best DRAM command to issue \circled{7} to serve a memory request, and 3) issuing the DRAM command \circled{8}, which updates the behavior and timing information of the DRAM device model. Finally, {the memory controller executes the finished} {request's callback} \circled{9} to notify the frontend {of the completion of the memory request}. 

Users can easily extend Ramulator 2.0 without intrusive changes to existing code by creating {different} implementations of {each} existing interface in three easy steps: 1) create a new \texttt{.cpp} file, 2) create the new implementation class that inherits from both the implementation base class and the existing interface class, and 3) implement the new {functionality} in the new implementation class. Similarly, a new interface can be added simply adding a \texttt{.h} file containing the abstract interface class definitions. All interfaces and implementations in Ramulator 2.0 \emph{register themselves} to a class registry that bookkeeps the relationship {among} different interfaces and implementations. {Using} this registry, Ramulator 2.0 automatically recognizes and instantiates different implementations for {each interface} from a human-readable configuration file. Users do \emph{not} need to manually maintain any boilerplate code to describe the relationships between interfaces and implementations.

\subsubsection{Memory Controller Plugins}
\label{sec:ctrl_plugin}
\begin{figure}[h]

\centering
  \includegraphics[width=0.97\linewidth]{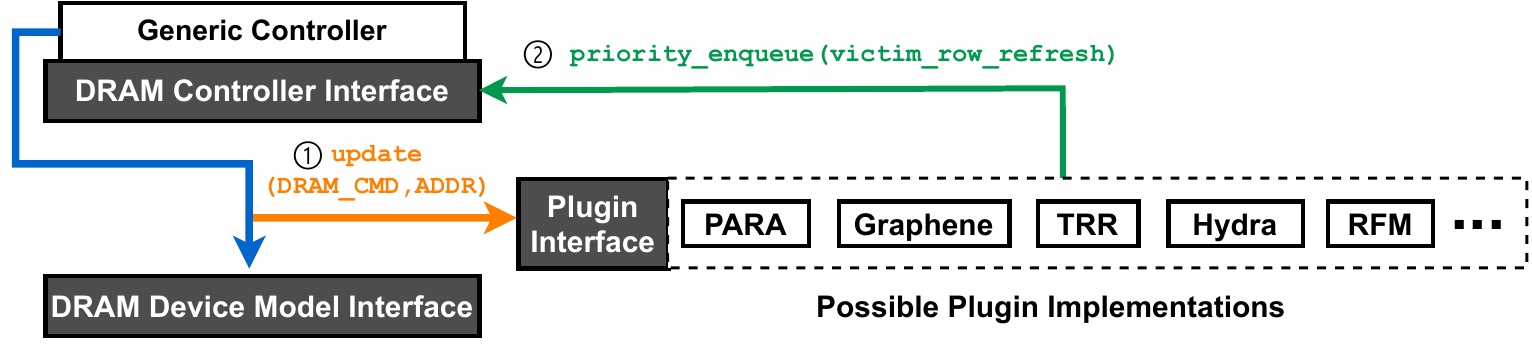}

  \caption{Implementing RowHammer mitigation {techniques} as controller plugins. {Legend is in Figure 1.}}
  \label{fig:plugin}
  \vspace{-1.5em}
\end{figure}

We make a key observation that many {modeled functions} in the memory controller (e.g., controller-based RowHammer mitigations that tracks the issued activation commands) and utilities needed for evaluation (e.g., collecting statistics from the issued DRAM commands and analyzing the memory access patterns) are {triggered (updated) by the currently-scheduled DRAM command.} To avoid having many similar memory controller implementations for every single such {modeled function and utility}, we model these {functions} as plugins to the memory controller. As an example, \figref{fig:plugin} shows in detail how various RowHammer mitigation {techniques} (e.g., PARA~\cite{kim2014flipping}, Graphene~\cite{park2020graphene}, Hydra~\cite{qureshi2022hydra}, {TRR~\cite{frigo2020trrespass, hassan2021utrr}}, RFM~\cite{jedec2020ddr5}) can be implemented as such {memory controller plugins.}

The plugin interface has a simple \texttt{update(DRAM\_CMD, ADDR)} function that the controller calls (\hcircled{1} in \figref{fig:ramulator_architecture} and~\ref{fig:plugin}) to notify the plugin implementations about the DRAM command and address issued by the memory controller. The RowHammer mitigation implementation then updates {its} internal state (e.g., generates a random number for PARA, {updates the row activation count table (bank activation counter) for Graphene and TRR (RFM)}, {or queries} the row count cache for Hydra). If the {implementation} detects the need to refresh the potential RowHammer victim rows, {it calls} the \texttt{priority\_enqueue()} function (\hcircled{2} in \figref{fig:ramulator_architecture}~and~\ref{fig:plugin}) of the memory controller interface to send a high-priority refresh request for the identified victim rows, ready to be scheduled in the {following} cycles, {as determined by the mitigation techniques}. To showcase the modularity and extensibility of memory controller plugins,~\secref{sec:rh} provides a cross-sectional evaluation of the performance overhead of six different RowHammer mitigation {techniques}, all implemented as memory controller plugins.

\subsection{Concise and Intuitive DRAM Specifications}
\label{sec:dram_model}
Ramulator 2.0 facilitates easy {modification} and {extension} of DRAM specifications {(e.g., the organization of the DRAM device hierarchy, DRAM commands, timing constraints, mapping between DRAM commands and organization levels)} in two major ways. First, Ramulator 2.0 allows the user to directly define the DRAM specifications \emph{by their names} with human-readable string literals, as {Listing} 1 shows.

\vspace{-.4em}
\begin{lstlisting}[style=cppstyle, caption={Example Definition of DRAM Organization and Commands}, label={lst:command_example}]
  // Different levels in the organizaton hierarchy
  inline static constexpr ImplDef m_levels = {
    "channel", "rank", "bankgroup", 
    "bank", "row", "column",    
  };
  // Different DRAM commands
  inline static constexpr ImplDef m_commands = {
    "ACT", "PRE", "PREab", "RD",  "WR", "REF"
  };
  // Mapping between commands and levels
  inline static const ImplLUT m_cmd_scopes = LUT (
    m_commands, m_levels, {
      {"ACT", "row"}, {"PRE", "bank"}, 
      {"RD",  "column"}, {"WR", "column"}, 
      {"REF", "rank"}, {"PREab","rank"},
    }
  );  
\end{lstlisting}
Internally, Ramulator 2.0 \emph{automatically} encodes these string literals into integers. {These integers are used to \emph{efficiently} index the the lookup table-based finite state machines that Ramulator 2.0 uses to model the hierarchical organization and behavior of DRAM devices, {similarly} to Ramulator 1.0~\cite{kim2016ramulator}}. This encoding is done statically at \emph{compile-time} within the frequently queried and updated DRAM device model so that it does \emph{not} incur any run-time performance overhead (e.g., the expression \texttt{m\_levels["bank"]} is a \texttt{consteval} expression that {is evaluated} to the integer ``3'' by the compiler). Other components in the simulated system that {need to know the DRAM device's specifications {(e.g., the organization of the DRAM device hierarchy, DRAM commands, timing constraints, the mapping between DRAM commands and organization levels)}} can query the DRAM specification with string literals during initialization to get the underlying integer encoding (or an error indicating the component is incompatible with the DRAM specification). Doing so \emph{completely} decouples the DRAM specifications from other parts of the simulated system, {thereby} achieving higher modularity and extensibility {than} Ramulator 1.0.

Based on these string-literal based definitions, Ramulator 2.0 develops a concise and human-readable {way to model} the timing constraints of DRAM commands. The key idea is to define timing constraints based on the permutation of the preceding and following DRAM commands. Doing so reduces redundant code by merging the timing constraint definitions that have the same numerical value but are between different pairs of preceding and following DRAM commands into a single definition. For example, Listing 2 shows the definition of the timing constraint \texttt{nRCD} that specifies the {minimum} delay between a preceding \texttt{ACT} command and either a following \texttt{RD} or \texttt{WR} command at the \texttt{bank} level. With this {modeling}, Ramulator 2.0 defines the key DDR4 timing constraints with only 32 lines of code, a 61\% reduction from Ramulator 1.0's 82 lines. {Such code deduplication enables the addition of new DRAM standards in an easier and less error-prone way.}
\vspace{-.4em}
\begin{lstlisting}[style=cppstyle, caption={Example Definition of Timing Constraints}, label={lst:command_example}]
  {.level = "bank", 
   .preceding = {"ACT"}, .following = {"RD", "WR"}, 
   .latency = V("nRCD")},  
\end{lstlisting}

Second, Ramulator 2.0 implements the {DRAM commands} (e.g., the state changes caused by the DRAM commands and the prerequisite commands based on the current state) using a library of lambda functions. These functions are implemented in a templated way so that they are {defined only once}, but can be \emph{reused} many times for similar DRAM commands across {\emph{different}} standards. {As} an example, Listing 3 shows a part of implementations of the \texttt{RFMab} command (all-bank refresh management, {which} exists in the DDR5~\cite{jedec2020ddr5}, LPDDR5~\cite{jedec2020lpddr5}, GDDR6~\cite{jedec2021gddr6}, and HBM3~\cite{jedec2023hbm3} standards) that requires all the banks to be closed before it can be issued.

\vspace{-.4em}
\begin{lstlisting}[style=cppstyle, caption={Example Implementation of a DRAM Command, \emph{RFMab} (shared across different DRAM standards, including DDR5, LPDDR5, GDDR6, HBM3)}, label={lst:command_example}]
  template <class DRAM_t>
  int RequireAllBanksClosed(typename DRAM_t::Node* node, int cmd, int target_id, Clk_t clk) {
    // for all banks {
    // ...
      if (bank->m_state == DRAM_t::m_states["Closed"]) {
        continue;
      } else {
        return T::m_commands["PREab"];
      }
    // }
    return cmd;
  };

  // In DDR5.cpp
  m_preqs[m_levels["rank"]][m_commands["RFMab"]] = RequireAllBanksClosed<DDR5>;
  // In LPDDR5.cpp
  m_preqs[m_levels["rank"]][m_commands["RFMab"]] = RequireAllBanksClosed<LPDDR5>;
  // In GDDR6.cpp
  m_preqs[m_levels["channel"]][m_commands["RFMab"]] = RequireAllBanksClosed<GDDR6>;
  // In HBM3.cpp
  m_preqs[m_levels["channel"]][m_commands["RFMab"]] = RequireAllBanksClosed<HBM3>;

\end{lstlisting}

{Ramulator 2.0 defines a \texttt{RequireAllBanksClosed} generic function} that checks for all banks in the organization hierarchy if all of them are closed ({lines 6-7}). If so, it simply returns the input command argument \texttt{cmd} (line 12), indicating {that} no prerequisite command is needed for \texttt{cmd}. Otherwise, it returns the \texttt{PREab} (precharge all-bank) command to close all the banks first. This function is templated on the DRAM standard implementation (i.e., the \texttt{DRAM\_t} template parameter {on} line 2) so that it can automatically get the correct integer encoding of the commands and states for different DRAM standards at compile-time. By reusing this templated function {in different DRAM standards (lines 16, 18, 20, and 22)}, implementing the prerequisite checks for the \texttt{RFMab} command needs only a single line of code in each standard {(instead of duplicating the entire \texttt{RequireAllBanksClosed} function {for each DRAM standard} as in Ramulator 1.0)}. 

\section{Validation \& Evaluation}
\subsection{Validating the Correctness of Ramulator 2.0}
To make sure Ramulator 2.0's memory controller and DRAM device model implementation is correct (i.e., the DRAM commands issued by the controller obey both the timing constraints and the state transition rules), we verify the DRAM command trace against Micron's DDR4 Verilog Model~\cite{micronddr4verilog} {using a similar methodology to prior works~\cite{kim2016ramulator,dramsim2,dramsim3}}. To do so, we implement a DRAM command {trace} recorder as a DRAM controller plugin that can store the issued DRAM commands with the addresses and time stamps using the DDR4 Verilog Model's format. We collect DRAM command traces from {eight} {streaming-access and eight random-access} synthetic memory traces and different intensities (i.e., the number of non-memory instructions between memory instructions). We feed the DRAM command trace to the Verilog Model, configured to use the same DRAM organization and timings as we use in Ramulator 2.0. {We find no timing or state transition violations.}
\subsection{Performance of Ramulator 2.0}
We compare the simulation speed of Ramulator 2.0 with three other cycle-accurate DRAM simulators: Ramulator 1.0~\cite{kim2016ramulator}, DRAMsim2~\cite{dramsim2}, DRAMsim3~\cite{dramsim3}, and USIMM~\cite{usimm}. All four simulators are {compiled with \texttt{gcc-12 -O3}, and} configured with comparable system parameters. We generate two memory traces, one with a random access pattern and another with a {streaming} access pattern, each {containing} five million memory requests (read-write ratio = 4:1). For each simulator {and trace}, we {run the simulation for each trace {ten} times on a machine with an Intel Xeon Gold 5118 processor.} Table~\ref{tab:perf-comp} shows the {minimum, average, and maximum simulation {runtimes} across the ten runs}. We conclude that, despite the increased modularity and extensibility, Ramulator 2.0 achieves a comparably fast {(and even faster)} simulation speed {versus} other existing cycle-accurate DRAM simulators. {We provide the scripts, configurations, and traces to reproduce {our} results in Ramulator 2.0's repository~\cite{ramulator2github}.}

\begin{table}[h]
\vspace{-.5em}

\centering
\resizebox{\columnwidth}{!}{%
\begin{tabular}{@{}ccccc@{}}
\toprule
\multirow{2}{*}{\textbf{\begin{tabular}[c]{@{}c@{}}Simulator\\ (gcc-12 -O3)\end{tabular}}} &
  \multicolumn{2}{c}{\textbf{\begin{tabular}[c]{@{}c@{}}Runtime (sec)\\ min./avg./max.\end{tabular}}} &
  \multicolumn{2}{c}{\textbf{Avg. Requests/sec}} \\ \cmidrule(l){2-5} 
                       & \textbf{Random}         & \textbf{Stream}         & \textbf{Random} & \textbf{Stream} \\ \midrule
\textbf{Ramulator 2.0} & \textbf{50.3/50.6/51.4} & \textbf{26.1/26.2/26.4} & \textbf{98.8K}  & \textbf{190.8K} \\
Ramulator 1.0          & 58.2/59.0/62.3          & 31.7/31.9/33.0          & 84.7K           & 156.7K          \\
DRAMsim3               & 51.4/51.7/52.3          & 37.5/37.8/38.6          & 96.7K           & 132.3K          \\
DRAMsim2               & 51.6/51.9/52.4          & 53.7/53.9/54.1          & 96.3K           & 92.8K           \\
USIMM                  & 402.9/407.0/410.0       & 31.2/31.3/31.4          & 12.3K           & 159.7K          \\ \bottomrule
\end{tabular}%
}
\vspace{-.5em}
\caption{Simulation Performance Comparison}
\vspace{-1.2em}
\label{tab:perf-comp}

\end{table}
\subsection{Cross-Sectional Study of RowHammer Mitigations}

\label{sec:rh}
\begin{figure}[h]
\vspace{-1.2em}
  \includegraphics[width=\linewidth]{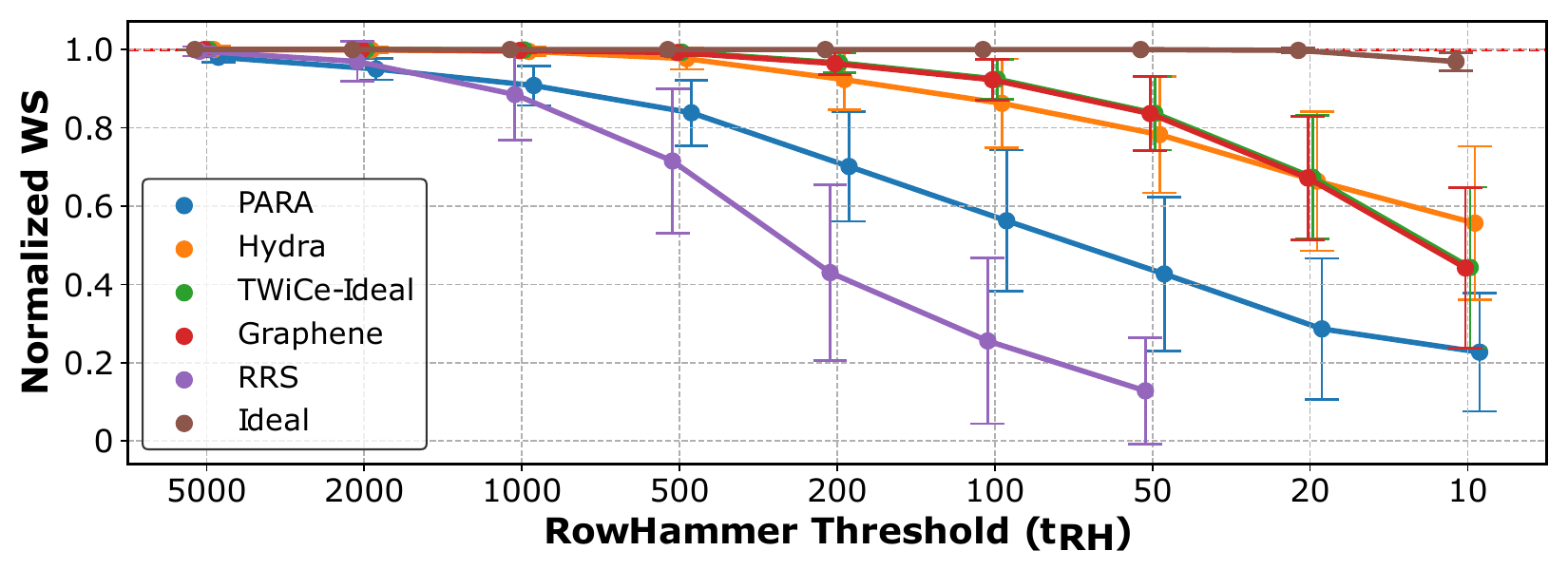}
  \vspace{-2.2em}

  \caption{Performance overhead of RowHammer mitigation {techniques vs.} different RowHammer thresholds}
  \label{fig:rh}
  \vspace{-1.3em}

\end{figure}

To demonstrate the modularity and extensibility of Ramulator 2.0, we implement six different RowHammer mitigation {techniques}, PARA~\cite{kim2014flipping}, an idealized version of TWiCe~\cite{lee2019twice}, Graphene~\cite{park2020graphene}, Hydra~\cite{qureshi2022hydra}, Randomized Row-Swap (RRS)~\cite{saileshwar2022randomized}, and an ideal refresh-based mitigation (Ideal)~\cite{kim2020revisiting}. {All of these mechanisms are implemented} in the form of {memory} controller plugins as described in~\secref{sec:ctrl_plugin}.
Figure~\ref{fig:rh} shows the performance overhead (weighted speedup normalized to a baseline configuration running the same workloads \emph{without} any RowHammer mitigation, y-axis) of different RowHammer mitigations as the RowHammer threshold (i.e., the minimum number of DRAM row activations to cause at least one bitflip, $\mathrm{t}_{\mathrm{RH}}$, x-axis) decreases from 5000 to 10. We use traces generated from SPEC2006~\cite{spec2006} and SPEC2017~\cite{spec2017} to form 25 four-core multiprogrammed workloads that we feed through a simplistic out-of-order core model {(the complete set of scripts and traces to reproduce these experiments are in~\cite{ramulator2github})}. 

{We make the following two observations. First, all evaluated RowHammer mitigations (except for Ideal) cause significant performance overhead compared to the ideal mitigation as $\mathrm{t}_{\mathrm{RH}}$ decreases to very low values. Second, for $\mathrm{t}_{\mathrm{RH}}$ $<50$, the performance overhead of RRS becomes too high for the simulation to make {progress}. {This is because} the activation caused by a row swap triggers even more row swaps, preventing DRAM from serving memory access requests. We conclude that existing RowHammer mitigation {techniques} are not scalable enough {to} {low $\mathrm{t}_{\mathrm{RH}}$ values ($<$50}). {As such,} more research effort is needed to develop more efficient and scalable RowHammer mitigation {techniques}.}

\section{Conclusion}
We present Ramulator 2.0, a modern, modular, and extensible DRAM simulator as a successor to Ramulator 1.0. We introduce the key design features of Ramulator 2.0 and demonstrate its high modularity, extensibility, {and performance}. We hope that Ramulator 2.0{'s modular and extensible software architecture and concise and intuitive modeling of DRAM} {facilitates} more agile memory systems research.

\ifCLASSOPTIONcaptionsoff
  \newpage
\fi

\bibliographystyle{IEEEtranS.bst}
\bibliography{refs}

\end{document}